\documentclass[traditabstract]{aa} 
\usepackage{graphicx}
\usepackage{natbib}

\usepackage{txfonts}
\begin{document}
   \title{``Comets'' orbiting a black hole}


   \author{R. Maiolino
          \inst{1}
          \and
          G. Risaliti
		  \inst{2,3}
		  \and
		  M. Salvati
		  \inst{2}
		  \and
		  P. Pietrini
		  \inst{4}
		  \and
		  G. Torricelli-Ciamponi
		  \inst{2}
		  \and
		  M. Elvis
		  \inst{3}
		  \and
		  G. Fabbiano
		  \inst{3}
		  \and
		  V. Braito
		  \inst{5,6}
		  \and
		  J. Reeves
		  \inst{7}
          }

   \institute{
  INAF-Osservatorio Astronomico di Roma, via di Frascati 33, 00040
              Monte Porzio Catone, Italy
	\and
	Osservatorio Astrofisico di Arcetri,
	Largo E. Fermi 5, 50125 Firenze, Italy
	\and
    Harvard-Smithsonian Center for Astrophysics,
	60 Garden St. Cambridge, MA 02138 USA   
	\and
	Dipartimento di Astronomia, Universit\`a di Firenze, largo E. Fermi 2,
	   50125 Firenze, Italy
	\and
	Department of Physics and Astronomy, University of Leicester,
	University Road, Leicester, LE1 7RH, UK
	\and
	Department of Physics and Astronomy, Johns Hopkins University,
	Baltimore, MD 21218
	\and
	Astrophysics Group, School of Physical and Geographical Science,
	Keele University, Keele, Staffordshire ST5 5BG, UK
             }

   \date{Received ; accepted }

 
  \abstract
   {
   We use a long (300~ksec), continuous {\it Suzaku} X-ray observation of the active
   nucleus in NGC~1365 to investigate the structure of the circumnuclear BLR clouds
   through their occultation of the X-ray source. The variations of the absorbing
   column density and of the covering factor indicate that the clouds surrounding
   the black hole are far from having a spherical geometry
(as sometimes assumed), instead they have a strongly elongated and cometary shape,
with a dense head ($\rm n\sim 10^{11}~cm^{-3}$)
and an expanding, dissolving tail. We infer that the cometary
tails must be longer than a few times $\rm 10^{13}~cm$ and their opening angle must
be smaller than a few degrees. We suggest that the cometary shape may be a common feature
of BLR clouds in general, but which has been difficult to recognize observationally
so far.  The cometary shape may originate from shocks and hydrodynamical instabilities generated by the
supersonic motion of the BLR clouds into the intracloud medium.
As a consequence of the mass loss into their tail,
we infer that the BLR clouds probably have a lifetime of only a few months, implying that
they must be continuously replenished. We also find a large, puzzling
discrepancy (two
orders of magnitude) between the mass of the BLR inferred from the properties of the absorbing clouds
and the mass of the BLR inferred from photoionization models; we discuss the possible solutions
to this discrepancy.
   }


   \maketitle
%

\section{Introduction}
\label{intro}

The variability of the absorbing gaseous column density $\rm N_H$
observed in the X-ray spectra of Active Galactic Nuclei have revealed that
a significant fraction of the absorbing medium must be clumpy
\citep{risaliti02}. Moreover, $\rm N_H$ variations on time scales
as short as a few days or hours have shown that a significant fraction of such
absorbing clouds must be located very close to the nuclear X-ray source
\citep{elvis04,risaliti05a,puccetti07} and, more specifically,
within the Broad Line Region
(BLR).

The nucleus of the galaxy NGC~1365 is amongst the sources that were
investigated more thoroughly in this respect, thanks to its
brightness and probably also to a fortunate (intermediate) inclination
of the absorbing medium relative to our line of sight. This is a nearby
(z=0.0055) type 1.8 Seyfert galaxy that was observed several times
in the X-rays, and displayed strong column density variations
on time scales as short
as a few days \citep[e.g.][]{risaliti05a,risaliti07}.
Most of the observations had probed 
absorption variations of this source in discrete, non-contiguous time
intervals. More recently, (nearly) continuous XMM observations of NGC~1365 were presented
in \cite{risaliti09a} and in \cite{risaliti09b}.
In particular, the latter observation allowed the continuous monitoring of
an absorption event lasting about 40~ksec, which was modelled with
a cloud partially eclipsing the X-ray source. More specifically,
the spectral variations were modelled by assuming a constant column
density of the absorber ($\rm N_H\sim 3.5~10^{23}~cm^{-2}$) and
a variable covering factor. The inferred density, velocity and distance from
the black hole of the variable absorber is strikingly similar to those
of the BLR clouds. The complex spectrum always requires
a foreground non-variable
(or very slowly variable) absorber ($\rm N_H\sim$ a few times
$10^{22}~cm^{-2}$), totally covering the source,
possibly associated with gas at larger distances from
the nucleus (possibly in the host galaxy).

NGC~1365 has also been observed with {\it Suzaku} \citep{mitsuda07} with a continuous observation
lasting more than 300~ksec. The integrated spectrum was already presented in
\cite{risaliti09c}, mostly focusing on the high energy part obtained with
the Hard X-ray Detector (HXD), and revealing an unusually strong excess of emission
at $\rm E>10~keV$. Such excess is ascribed to X-ray radiation piercing
a partial, Compton thick ($\rm N_H \sim 4\times 10^{24}~cm^{-2}$)
absorber. The high energy excess appears stable over long time
scales \citep{risaliti00,risaliti09c}. The emerging scenario
is that three distinct absorbers are present: (1) a distant
absorber with $\rm N_H < 10^{23}~cm^{-2}$, probably associated with gas in the
host galaxy; (2) an absorber
made of broad line region (BLR) clouds with $\rm N_H\sim 10^{23-24}
cm^{-2}$ rapidly orbiting around the black hole; (3) an absorber with
$\rm N_H \sim 4\times 10^{24}~cm^{-2}$ partially covering the source,
consisting either of the outer region of
a warped accretion disk, or of a large number of small ($\rm <10^{12}cm$) and
dense ($\rm n>10^{12}cm^{-3}$) clouds located in the vicinity of the accretion
disk.

In this paper we focus on the absorption variations
in the spectral range 2-10~keV, ascribed to BLR clouds passing along the
line of sight, through a detailed 
temporal analysis of the spectra provided by the {\it Suzaku} low energy (XIS)
data of NGC~1365.
We identify two eclipsing
events that we can model in detail in terms of variations of both covering factor
and column density. The time variations of these quantities are
highly asymmetric and strongly suggest that the absorbing clouds have
a cometary shape, i.e. a well defined head and a more diffuse elongated,
conical, tail.

\section{Observations and data reduction}
\label{observations}

The observations and the basic data reduction were already reported
in \cite{risaliti09c}. In this section
we only briefly summarize the observations
and data reduction of the X-ray Imaging Spectrometer \citep[XIS][]{koyama07} data.

The observations were performed on 2007 January 21-25, for an elapsed
time of about 320~ksec
and a net exposure of 160~ksec.
The spectra and calibrations were obtained following the standard procedure
described in the {\it Suzaku} reduction guide. The XIS spectra and light
curves were extracted
in a region with a radius of 2.9 arcmin (i.e. matching the optimized aperture
for the arf matrix) around the nucleus of NGC~1365.
The background was extracted in a region far from the source, with
the same radius adopted for the source extraction.
The calibration files were produced by using the FTOOLS 6.6 package.
The reduction of the Hard X-ray Detection (HXD) data, only marginally
used in this paper, is discussed in \cite{risaliti09c}.

  \begin{figure}
   \centering
   \includegraphics[angle=0,width=9truecm]{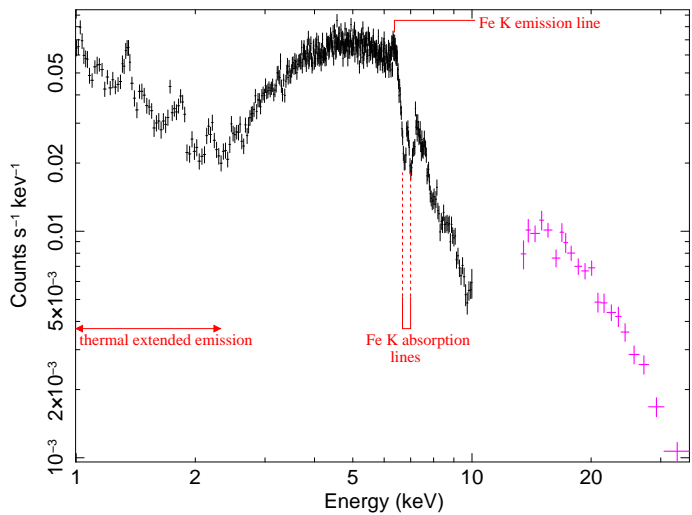}
      \caption{
         \label{spec_tot}
	  Merged XIS0 and XIS3 spectra (1--10 keV) and HXD spectrum (15--60~keV)
	  of NGC~1365,
	  integrated over the whole duration of the observation.
	  No spectral fit is shown since, as discussed
	  in the text, the absorption components are highly variable,
	  hence a fit (with constant components) to the total, time-averaged spectrum
	  is meaningless.
              }
   \end{figure}

\section{Temporal and spectral analysis}
\label{analysis}

In this paper the spectra
from the three CCDs (XIS0, XIS1 and XIS3) are always treated separately,
especially in the statistical analysis. However, for sake of clarity,
some of the figures will show the merged spectra or light curves,
for illustrative purposes only.

In the 0.5--10~keV
range during the whole observation the three XIS collected a total
of about 2~10$^5$ photons. Fig.~\ref{spec_tot} shows the XIS0 and XIS3
merged spectrum integrated over the whole observation along with the HXD spectrum.
No spectral fit is shown
for the total, time-averaged spectrum, since the strong spectral variability (especially of
the absorption components) over the
entire observation, as discussed in the following, makes any fit with constant
components of the integrated
spectrum meaningless \citep[as also discussed in ][]{risaliti09b}.

\begin{figure}[!]
   \centering
   \includegraphics[bb=57 82 550 777,angle=0,width=9truecm]{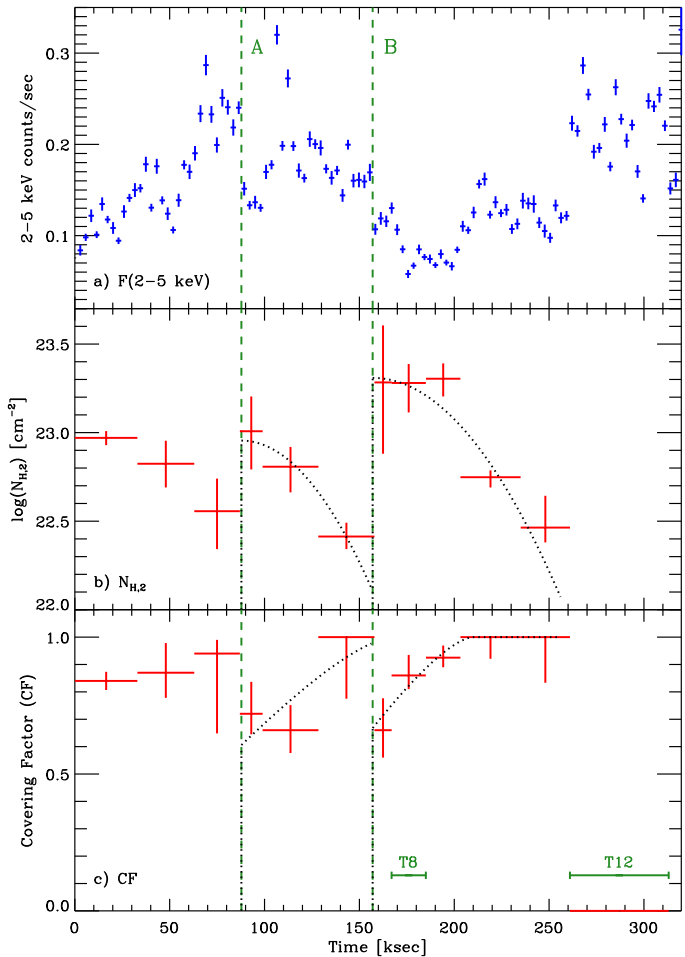}
      \caption{
         \label{eclipses}
	  a) Photon count rate in the soft band 2-5 keV as a function of time
	  (in time intervals 2.88 ksec).
The vertical, green dashed lines indicate the beginning of the two main eclipses produced
 by two clouds passing in front of the X-ray source. The vertical errorbars indicate
 the 1$\sigma$ fluctuations within each bin. 
	  b) Column density ($\rm N_{H,2}$) of the variable absorber in the 12 time
	  intervals, resulting from their simultaneous spectral
	  fit. Errorbars on $\rm N_{H,2}$ are at the 90\% confidence.
	  The black dotted lines indicate a quadratic fit to $\rm N_{H,2}$ in
	  the eclipses ``A'' and ``B''.
	  c) Covering factor ($\rm CF$) of the variable absorber.
	 Errorbars on $\rm CF$ are at the 90\% confidence.
	  The dotted lines show the 
	  variation of covering factor expected from the cometary model shown in Fig.~\ref{model}
	  and discussed in the text.
	  The segments labelled with T8 and T12 show the extraction time intervals of the
	  spectra shown in Fig.~\ref{spec_ind}.
              }
   \end{figure}

Fig.~\ref{eclipses}a shows the combined light curve in the 2--5~keV band
rebinned to 2.880 ksec (about half of the {\it Suzaku} orbital period).
The light curve shows strong variations even on short timescales of a few ksec,
including two sharp drops (marked A and B in Fig.~\ref{eclipses}) that are
strongly reminiscent of the eclipse events seen in Chandra and XMM. The egress
from these ``eclipses'' is less clear cut however, suggesting more complex structure in
the absorber.

\begin{figure}
   \centering
   \includegraphics[angle=-90,width=8truecm]{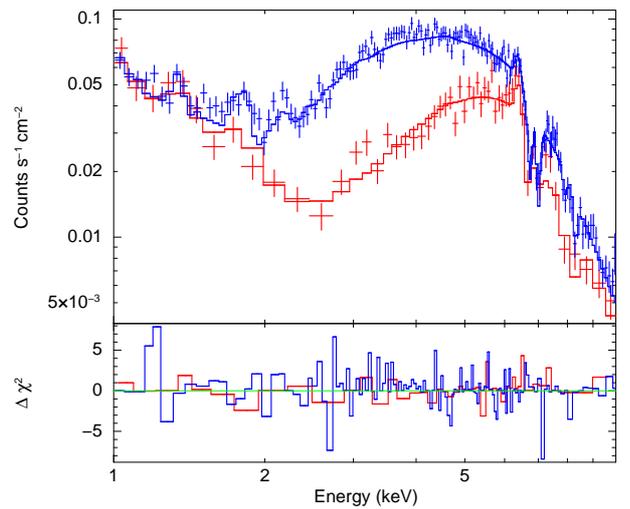}
      \caption{
         \label{spec_ind}
	  Merged XIS0 and XIS3 spectra of NGC~1365 and resulting fitting model in two representative
	  time intervals, T8 (between 167 ksec and 185 ksec, red) and T12
	  (between 261 ksec and 313 ksec, blue), as indicated in Fig.~\ref{eclipses}.
	  Although, for the sake of clarity,
	  we plot the combined XIS0 and XIS3 spectra in both figures,
	  in the fitting analysis we used all individual XIS spectra (XIS0,
	  XIS1 and XIS3), kept separate.
              }
   \end{figure}

To investigate the nature of these variations we have extracted the X-ray
spectra in 12 time intervals, selected either
to match the main variations of the light curve 
or by simply dividing long time intervals
 into equally spaced sub-intervals.
 Within each time interval the light curve generally shows additional structures,
 but we decided not to further split the time intervals for the extraction of
 the spectra, since the resulting poorer statistics would make the errors on
 the fitting parameters too large, or would introduce strong degeneracy between some
 of them.
 On average, each time interval contains
 about 1.5~10$^4$ photons (e.g. Fig.~\ref{spec_ind}).

 The resulting 36 spectra (for
 each time interval there are 3 spectra, XIS0, XIS1 and XIS3)
 are simultaneously fitted with a model
 consisting of multiple components,
 quite similar to previous modelling of the same source \citep{risaliti09b,guainazzi09},
 with most 
 components and parameters held constant for all time intervals (and hence constrained by
 the full statistics of 2~10$^5$ photons) and a few parameters left free to vary for
 the different time intervals.
 In the following we summarize the various models components.

\begin{enumerate}
\item {\it Soft thermal components.} As reviewed and discussed in \cite{guainazzi09} the soft part of the
 spectrum (0.5-2 keV) is ascribed to a complex combination of hot plasmas present
 in the circumnuclear star forming regions \citep{wang09} and in the Narrow Line Region cones,
 as well as the contribution from point sources \citep[X-ray binaries and Ultra
 Luminous X-ray sources, ][]{soria09}. Following the detailed analysis in \cite{guainazzi09}
 we fit the soft spectrum with three components: two absorbed thermal plasmas
 (``mekal'' in XSpec)
 and a black-body. The resulting best fit parameters are generally in agreement
 with those obtained by \cite{guainazzi09} (see below), keeping in mind that
 the aperture adopted here is different,
 hence we do not expect an exact match.
However, the details of these soft components are unimportant within
the context of this paper,
since they are constant in time and do not affect the
variability observed in the 2-10 keV spectral region.

\item {\it Powerlaw and two cold absorbers.} The bulk of the 2-10~keV emission is due to 
 a primary powerlaw component, ascribed to the corona surrounding
 the black hole accretion disk,
 with a double absorber: one absorber ($\rm N_{H,1}$)
 totally covering the X-ray source, which we shall ascribe to
 a foreground medium far from the X-ray source, and a second absorber ($\rm N_{H,2}$)
 partially covering the source, which, following \cite{risaliti09b},
 we shall ascribe to clouds transiting close to
 the X-ray source (probably BLR clouds). By using wider time intervals ($\sim$50~ksec)
 we found that the spectral index does not change within the errors 
 \citep[as also found by ][]{risaliti09b}. Therefore, we assumed that
 the spectral index is the same for all 12 time
 intervals. The powerlaw normalization is left free
 to vary among the time intervals. The first, total, absorber is kept constant for all
 time intervals.
 The absorber with partial
 covering is left free to vary both in terms of column density ($\rm N_{H,2}$)
 and in terms of covering factor ($\rm CF$).

\item
 {\it Narrow FeK line}.
 A narrow (unresolved) iron FeK$\alpha$ line at $\sim$6.4 keV is clearly
 detected. In larger temporal bins the flux of this line is not observed to
 vary. As discussed by various previous works the narrow FeK line is probably produced by
circumnuclear
 dense clouds (most likely Broad Line Clouds and circumnuclear absorbing medium in general).
 The parameters
 of this line are therefore kept constant and fitted simultaneously for all time intervals.
 
\item
 {\it Broad FeK line}.
 As discussed in \cite{risaliti09b} the spectrum of NGC~1365 (when not
 in a Compton thick state) also
 requires a broad, relativistic iron FeK$\alpha$ line. 
 The parameters of the iron line (fitted with the ``laor'' function
 within Xspec) are kept constant for all time intervals, except for its
 normalization that is required to be proportional to the primary powerlaw in
 each time interval (i.e. constant equivalent width).
 The physical justification of this assumption is that,
 according to standard
 models, the relativistically broadened iron line comes from a very small region,
 of the order of a few Schwarzschild radii,
\citep[$\rm M_{BH}\sim 2~10^6~M_{\sun}$, hence $\rm R_S\sim 6~10^{11}~cm$, ][]{risaliti09b}
 and therefore it should respond to the
 continuum variations on timescales much shorter ($\sim$100 sec) than the size of our time
 intervals. However, in a few representative intervals we have tried to leave the
 normalization of the broad line free, and the fitting
 results do not change significantly, the main effect being mostly that
 the confidence intervals increase by about 20--30\%.
 
\item
{\it FeXXV and FeXXVI absorption lines.}
 As first discovered by \cite{risaliti05b} the spectrum of NGC~1365 is
 characterized by
  a set of four absorption lines at $\sim$7~keV due to Fe XXV
 and Fe XXVI, both K$\alpha$ and K$\beta$. These lines are also clearly detected
 in our {\it Suzaku} spectrum (Fig.~\ref{spec_tot}).
 As in \cite{risaliti05b}, the
 depth of each absorption line is left free to vary independently in each time
 interval, but the four lines are required to have the same velocity shift within each
 interval. We note that given the high ionization state of this absorber, its
 only spectral components that we expect to be important are just the Fe absorption lines,
 i.e. there is no significant continuum photoelectric absorption associated with this
 absorber.
 
\item
 {\it Reflection components.}
 As in \cite{risaliti09b} the fit also includes
 a set of three (constant in time) continuum
 reflection components: i) a power law with the same
 spectral index as the primary continuum, which accounts for the free electron scattering
 by an ionized reflecting
 medium on large scale (e.g. hot gas in the Narrow Line Region), hence this component
 is assumed constant over the whole observation; ii)
 a cold reflection (modelled with ``pexrav'' in Xspec), associated with the same
 medium responsible for the FeK$\alpha$
 narrow line emission, likely the circumnuclear
 dense clouds,
 hence it is assumed to be constant, as in the case of the narrow FeK;
 iii) an inner ionized absorber (modelled with ``pexriv'' in
 Xspec and subject to the same absorption as the primary power law component) to
 account for warm reflection due to ionized gas close to the X-ray source,
possibly the accretion disk itself; this third component is assumed, as for the
broad FeK line, to respond to the continuum variations on timescales much shorter
than our time intervals,
 hence its normalization  is assumed to scale proportionally to the
 primary powerlaw
 continuum within each time interval. We will see that the reflection components i) and iii)
 contribute very little to the global fit. Finally, we note that component iii) may be relativistically
 blurred, however given its small contribution to the spectrum such ``blurring'' cannot be
 constrained and it does not affect the total fit.

\item
 {\it Compton thick absorbed powerlaw.}
 As mentioned in sect.~\ref{intro}, the {\it Suzaku} spectrum at
 $\rm E>15 keV$ (HXD) shows an excess of 
 emission with respect to the extrapolation of the above components \cite{risaliti09c}.
 This excess is best interpreted by assuming that the intrinsic X-ray emission is
 actually stronger than observed in the 2-10 keV spectrum,
 but partially absorbed by gas very close to the X-ray source
 that we fit with $\rm N_{H,3}=3.7_{-1.0}^{+1.4}~10^{24}~cm^{-2}$. The inferred covering factor
 strongly depends on the geometry of the Compton thick medium, as discussed
 in \cite{risaliti09c}.
 Compton thick partial covering has also
 been observed in other sources \citep{turner09}. This inner,
 very thick, partial absorber is either associated with gas in a warped
 accretion disk or a population of small and dense clouds surrounding
 the accretion disk. The very
 high-energy excess is found to be stable over long time scales (years), hence in our
 analysis of the {\it Suzaku} data this component was assumed constant for all time
 intervals. We note that this additional component at high energies does not
 affect the fitting of $\rm N_{H,2}$ and $\rm CF$.

\end{enumerate}

The global, simultaneous fit gives a reduced $\chi^2_{\nu}$=1.055.
Overall there are 25 free parameters common to all time intervals that,
therefore, are constrained by the full 2~10$^5$ photon statistics. Each time interval
has its own set of 8 free parameters, three for the continuum ($\rm N_{H,2}$, $\rm CF$ and
powerlaw normalization) and five for the absorption Fe lines (velocity and individual depth).

 Fig.~\ref{spec_ind} shows the
 spectra and the resulting fit in two representative time intervals
 (T8 and T12, as marked in
 Fig.~\ref{eclipses}), clearly
 displaying a strong variation of absorption between the two time intervals.

 In the following we will mostly focus on the variations of the column density
 $\rm N_{H,2}$ and of the Covering Factor ($\rm CF$) of the partial absorber,
 whose resulting best-fit values in the various time intervals are shown in
 Fig.~\ref{eclipses}b and c.

\begin{figure}[!]
   \centering
   \includegraphics[angle=0,width=8truecm]{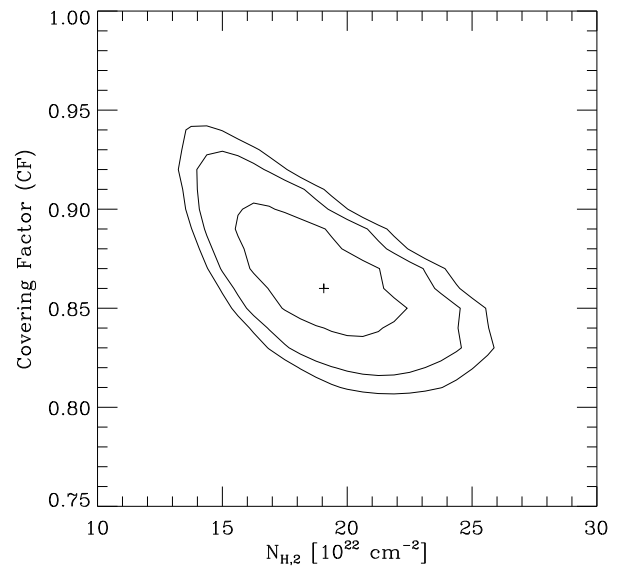}
      \caption{
         \label{contour}
	68\%, 90\% and 95\% confidence levels 
	  in the $\rm N_{H,2}$ versus Covering Factor
	  diagram, for the time interval T8 (Fig.~\ref{eclipses}).
	  The cross marks the best fit parameters.
              }
   \end{figure}

 One possible concern is whether
 these two quantities are degenerate, given the reduced statistics available
 in each time interval. We found this not to be the case in any of the time
 intervals (except for the last interval, which does not require a second absorber).
 As an example, Fig.~\ref{contour} shows the confidence levels in
 the $\rm N_{H,2}$ versus $\rm CF$ plane, for the time interval T8:
 although there is some correlation between the two parameters
 around the best solution,
the 90\% confidence contour provides good constraints on both quantities.
We emphasize that this is the first time that a temporally
 resolved X-ray spectral analysis is able to break the degeneracy between the
 evolution of the column density ($\rm N_H$) and the covering factor (CF) of the X-ray
 absorber in an AGN.

In the rest of this section we report the fitting results for the other parameters,
which, although important for the global fit, are less relevant for the topic if this paper.

Table~\ref{pars_comm} lists the best fitting values for the main parameters in common
to all time intervals.
The first part reports the soft (not variable) thermal components.
The second part reports the powerlaw slope and the reflection components.
Note that the continuum slope $\Gamma$ obtained here is different than obtained
by \cite{risaliti09c} on the same set of data, i.e. $\Gamma =2.34^{+0.03}_{-0.02}$. There
are two clear $\chi^2$ minima for this parameter, but the detailed analysis in
sub-intervals performed here (hence accounting for the spectral variability) shows
that the $\chi^2$ minimum with $\Gamma =1.81$ fits the data with much higher significance
($\chi_{\nu}^2=1.05$, versus ${\chi}_{\nu}^2=1.08$, with about a thousand d.o.f).
The reflection factor R of the reflection
components is given relative to the 2--10~powerlaw continuum observed in the last
time interval, T12
(Fig.~\ref{eclipses}). The real value of the reflection factor should be given relative to the
intrinsic powerlaw component, i.e. the continuum observed at E$>$20~keV corrected for
the Compton thick partial covering $\rm N_{H,3}$, which is much higher than the component
observed in the 2--10~keV range and which gives values of R that are much lower.
However, applying the correction to the continuum transmitted at high energies is very
uncertain, since it depends strongly on the geometry of the Compton thick absorbing
medium, as discussed in detail in \cite{risaliti09c}.

The third part of Table~\ref{pars_comm} reports the constant absorption components.
For what concerns the Compton thick absorbing medium, $\rm N_{H,3}$ only accounts
for the photoelectric absorption, which is enough for our purposes, since we are not
interested in the high-energy component. The full treatment including the Compton
scattering is given in \cite{risaliti09c}.

The fourth part of Table~\ref{pars_comm} provides the parameters of the emission lines.
The equivalent widths are given relative to the last time interval, T12.

\begin{table}[!h]
\caption{Best fit parameters common to all time intervals}
\label{pars_comm}
{\centering
\begin{tabular}{ll}
\hline\hline                 
\multicolumn{2}{c}{Soft thermal components} \\
 & \\
$\rm kT_1$ (mekal) & $\rm 0.30^{+0.04}_{-0.03}~keV$\\
$\rm N_H^{T1}$ & $\rm 1.88^{+0.8}_{-0.7}~10^{22}~cm^{-2}$ \\
$\rm kT_2$ (mekal) & $\rm 0.66^{+0.03}_{-0.03}~keV$\\
$\rm N_H^{T2}$ & $\rm 0.24^{+0.1}_{-0.1}~10^{22}~cm^{-2}$ \\
$\rm kT_3$ (black-body) & $\rm 0.16^{+0.02}_{-0.02}~keV$\\
 & \\
\hline
\multicolumn{2}{c}{Powerlaw and reflection components} \\
 & \\
$\Gamma$ & $\rm 1.81^{+0.03}_{-0.06}$ \\
$\rm R_{COLD}$ & $\rm 2.6^{+1.0}_{-1.0} $\\
$\rm R_{ION}^{in}$ & $\rm 0.22^{+0.20}_{-0.10} $\\
$\rm R_{ION}^{out}$ & $\rm 0.05^{+0.05}_{-0.03} $\\
 & \\
\hline
\multicolumn{2}{c}{Constant absorption components} \\
 & \\
$\rm N_{H,1}$ & $\rm 7.5^{+0.3}_{-0.7} 10^{22}~ cm^{-2}$\\
$\rm N_{H,3}$ & $\rm 3.7^{+1.4}_{-1.0}~10^{24}~ cm^{-2}$\\
\hline
\multicolumn{2}{c}{Emission lines} \\
 & \\
$\rm E_N^a$ & $\rm 6.37^{+0.03}_{-0.03}  keV$\\
$\rm F_N^a$ & $\rm 1.25^{+0.1}_{-0.1}  10^{-5} keV~cm^{-2}~s^{-1}~keV^{-1}$\\
$\rm EW_N^a$ & $\rm  52^{+4}_{-4} eV$\\
$\rm E_B^b$ & $\rm 6.86^{+0.04}_{-0.04}  keV$\\
$\rm F_B^b$ & $\rm 1.2^{+0.1}_{-0.1}  10^{-4} keV~cm^{-2}~s^{-1}~keV^{-1}$\\
$\rm EW_B^b$ & $\rm  390^{+40}_{-40} eV$\\
$\rm R_{IN}$ & $\rm  2.6^{+0.8}_{-0.8} (R_G)$\\
$\rm R_{OUT}$ & $\rm  7.1^{+3.0}_{-2.6} (R_G)$\\
$\rm \theta ^c$ & $\rm  25^{+15}_{-10} deg$\\
$\rm q^d$ & $\rm  2.9^{+0.9}_{-0.9}$\\
 & \\
\hline\hline
\end{tabular}
}
\\
Notes:\\
$\rm ^a$ Energy, flux and equivalent width of the narrow component of the FeK$\alpha$ line.\\
$\rm ^b$ Energy, flux and equivalent width of the broad component of the FeK$\alpha$ line.\\
$\rm ^c$ Disk inclination angle in the broad line component.\\
$\rm ^d$ Disk emissivity index in the broad line component.
\end{table}


   \begin{figure}[!]
   \centering
   \includegraphics[angle=0,width=8truecm]{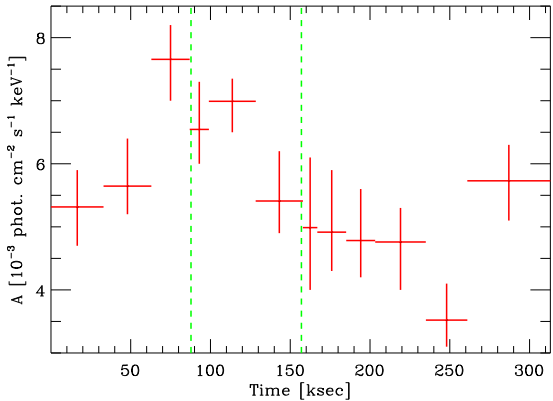}
      \caption{
         \label{norm}
	Variation of the normalization of the powerlaw continuum in the 2--10~keV range over
	the 12 time intervals analyzed in Sect.~\ref{analysis}. The green dashed lines
	indicate the beginning of eclipses A and B, as in Fig.~\ref{eclipses}.
              }
   \end{figure}

Fig.~\ref{norm} shows the variation of the normalization of the powerlaw continuum
in the 2--10~keV range in the same intervals analyzed in Fig.~\ref{eclipses}.
The normalization changes significantly during the observations. This is not unexpected
since NGC~1365 is a Narrow Line Sy1 and, as such, it is expected to be characterized
by significant variability. It is important to note that the variations in
the normalization of the powerlaw are uncorrelated with the 2-5~keV light curve behavior,
further supporting the idea that the variations of the light curve also require variation of the
absorbing column density and covering factor.

   \begin{figure}[!]
   \centering
   \includegraphics[angle=0,width=8truecm]{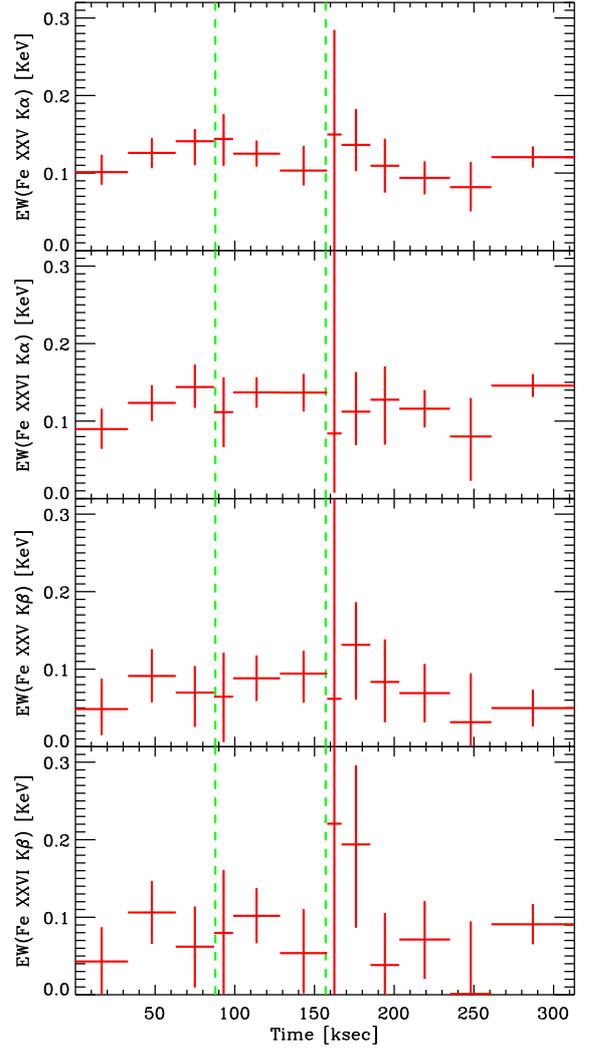}
      \caption{
         \label{abs_lines_ew}
	Equivalent width of the Fe absorption lines over
	the 12 time intervals analyzed in Sect.~\ref{analysis}. The green dashed lines
	indicate the beginning of eclipses A and B, as in Fig.~\ref{eclipses}.
              }
   \end{figure}

   \begin{figure}[!]
   \centering
   \includegraphics[angle=0,width=8truecm]{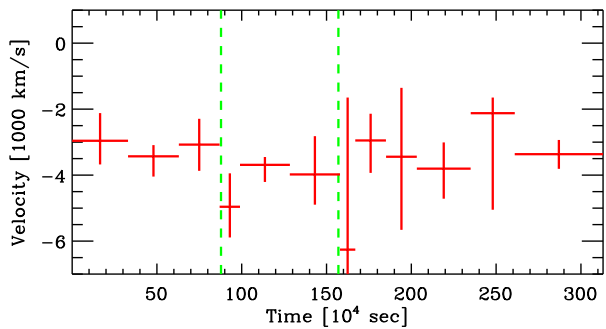}
      \caption{
         \label{abs_lines_vel}
	Velocity of the Fe absorption lines over
	the 12 time intervals analyzed in Sect.~\ref{analysis}. The green dashed lines
	indicate the beginning of eclipses A and B, as in Fig.~\ref{eclipses}.
              }
   \end{figure}

Fig.~\ref{abs_lines_ew} shows the equivalent width of the four Fe absorption lines
(FeXXV K$\alpha$ 6.697~keV, FeXXVI K$\alpha$ 6.966~keV, FeXXV K$\beta$~7.880,
FeXXVI K$\beta$ 8.268~keV) in the 12 time intervals analyzed in Sect.~\ref{analysis}.
Although there is some indication of variation of the depth
of the absorption lines, especially in association with eclipse B, the variations are all within
the statistical errors. Fig.~\ref{abs_lines_vel} shows the bulk velocity of the absorption lines,
which also does not show significant changes within errors. We note that the inferred EW are in 
line with those obtained for the same source by \cite{risaliti05b}. In the latter work variability
of the lines (both in terms of EW and of velocity) was observed on timescales of a few months.
However, our data are not sensitive to small changes in the absorption lines EW and velocity
shift on much shorter timescales, with lower exposures.

\section{The cometary shape of the eclipsing clouds}
\label{comets}

From Fig.~\ref{eclipses}, and in particular from the light curve and
the $\rm N_{H,2}$ variations,
we identify two main eclipsing events, labelled ``A'' and ``B'',
due to the passage of two clouds in front of the X-ray source,
with eclipse ``B'' being the more prominent and more clear event.
The decreasing $\rm N_{H,2}$ in the time interval prior to eclipse ``A'' probably shows the final
part of the occultation by a previous cloud.
The eclipses are far from being symmetric, implying that clouds are not spherical.
A spherical cloud significantly larger than the X-ray source would either produce
a sudden total obscuration ($\rm CF=1$)
or a smooth and symmetrical evolution of the covering factor if
grazing the X-ray source; while a spherical cloud with size comparable to, or smaller than the
X-ray source would simply produce a short, symmetric dip in the light curve and in the
covering factor.

Instead, both eclipses start with an abrupt drop of the 2--5~keV flux (marked by
the green vertical dashed lines A and B in Fig.~\ref{eclipses}) occurring within less
than a few thousand seconds. The spectral analysis of the first time interval
of each eclipse (Fig.~\ref{eclipses}b and c) indicates that such a sudden drop
of X-ray flux is due to the transit of a
cloud with $\rm N_H\sim 10^{23}~cm^{-2}$ suddenly
covering about 65\% of the source. Subsequently the
covering factor increases, but more slowly, on time scales of about 50 ksec.

These observational results indicate that the occulting cloud must be elongated, with a
size perpendicular to the direction of motion comparable to the X-ray source, while
the size along the direction of motion is at least an order of magnitude larger.
This is an immediate consequence of the fact that the beginning of the eclipse is
very sharp (with the covering factor jumping to 65\% in about a ksec) while the
subsequent evolution of the covering factor is much smoother on timescales of several 10 ksec.

The fact that the covering factor slowly increases, reaching unity about
50~ksec after the beginning of the eclipse, indicates not only that the clouds
are elongated, but that they must have a conical,
cometary shape, with their tail gradually occulting the whole X-ray source.

The decrease of
column density with time within each eclipse indicates that the head of
the ``comet'' is much denser than its tail, while the latter fades gradually into
a low density medium. We note that even the time interval before eclipse ``A'' shows a
similar trend, i.e. decreasing $\rm N_{H,2}$ and increasing covering factor,
hence in these intervals we may be observing the tail of a comet whose
head passed before the beginning of the observation.

We can quantify the geometry of these ``cometary'' clouds, as discussed in the
coming sections.
Note however, that
their basic scale depends heavily on the clouds' velocities
and on the dimension of the X-ray source, which are both unknown, although some
constraints can be set.

\begin{figure*}[!]
   \centering
   \includegraphics[angle=-90,width=18truecm]{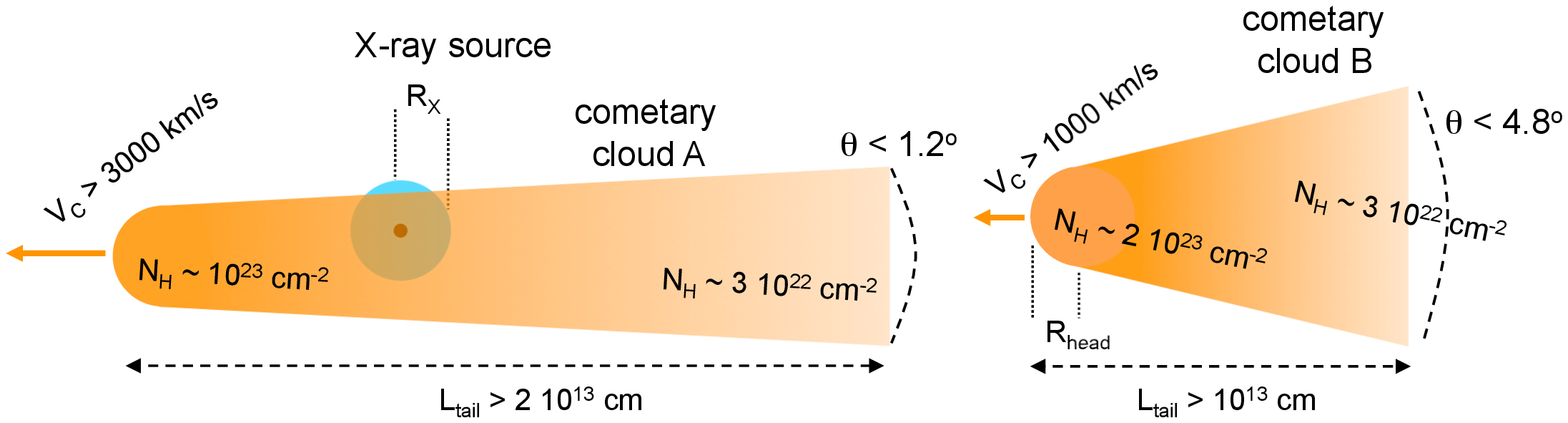}
      \caption{
         \label{model}
	  Sketch illustrating the geometry of the cometary clouds that eclipse the X-ray source,
	  as inferred by our observations. For the sake of clarity the drawing dimensions are not 
	  in the correct scale (in particular the horizontal length of the tails should be much
	  longer when compared to the size of the cloud heads and the tail opening angles are
	  exaggerated). The variation of the covering factor expected by this model is
	  shown with a dotted line in Fig.~\ref{eclipses}c.
              }
   \end{figure*}

The sketch in Fig.~\ref{model} summarizes the geometrical constraints inferred
for the two clouds derived below.
This is obtained under the assumption that the tail is elongated exactly along the direction
of motion of the cloud.
Moreover, based on the analysis of the temporal variation
of the covering factor we can only reconstruct the shape of the cometary
tail {\it projected} on the plane of the sky.
Our data cannot probe any elongation of the tail along the line of sight.

In principle the observed variations can be interpreted with other, more
complex geometries
(e.g. curved clouds with density gradients along the direction of motion), but the
cometary shape is the simplest
model that can account for both the variations of covering factor and of column
density, and which can be interpreted in terms of physically plausible scenarios,
as is discussed later on in sect.~\ref{nat}.

\subsection{The clouds' velocity}
\label{velocity}

If we assume that the cloud head is a hemisphere (Fig.~\ref{model}), then
the fact that the initial covering factor of each eclipse is large ($\sim$65\%), but
not total, along with the fact that the initial flux drop is very sharp, indicate
that the cometary cloud head and the X-ray source have similar sizes.
The sharpness of the flux drop at the beginning of each eclipse provides constraints
on the ratio $\rm R_X/V_C$ between the size of the X-ray source ($\sim$ size of the cloud head),
and the cloud velocity (in the plane of the sky).

\begin{figure}[!]
   \centering
   \includegraphics[angle=0,width=8truecm]{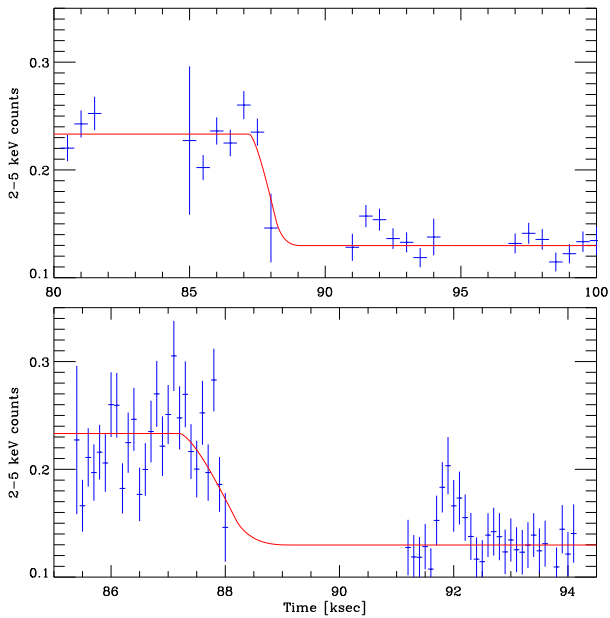}
      \caption{
         \label{init_eclA}
	  Detail of the 2--5~keV light curve around the beginning of eclipse ``A''.
	  The top panel is with bins of 500~sec. The bottom panel is
	  with bins of 100~sec and the time axis is further expanded around the
	  beginning of eclipse. The red solid line indicates the
	  modelled light curve where the
	  ratio between the size of the X-ray source ($\sim$radius of the
	  cloud head)
	  and the cloud velocity is $\rm R_X/V_C=10^8~cm~(km/s)^{-1}$ (see text).
	  If $\rm R_X= R_S$ this implies $\rm V_C=6000~km/s$.
              }
   \end{figure}

Fig.~\ref{init_eclA} shows the detail of the 2--5~keV light curve around the
beginning of eclipse ``A'', with a binning of 500~sec (top) and 100~sec (bottom).
The gaps are due to the Earth occultation during the {\it Suzaku} orbit. The sudden flux
drop due to the passage of the head of the cloud is marginally resolved,
just before the
Earth occultation gap, and it lasts about one ksec, or less.
We modelled the expected light
curve with a simple model where the X-ray source has a radius $\rm R_X$, the
cometary cloud head is a hemisphere with radius $\rm R_{head} \sim R_X$, moving
with a velocity $\rm V_C$ transverse to the line of sight, and covering 65\% of the
X-ray source right after the ``head'' transit. The minimum
possible radius for the X-ray source is given by the minimum stable orbit in the
case of a maximally rotating black hole, which, in the case of NGC~1365
implies $\rm R_X>0.5~R_S=3~10^{11}~cm$ (sect.~\ref{analysis}).
This, very conservative, lower limit on the size of the
X-ray source, along with the short timescale of the beginning of the eclipse
($\rm \le1ksec$),
sets a lower limit on the velocity of cloud ``A'', $\rm V_C(A) > 3000~km/s$.
By using these
lower limits (i.e. $\rm R_X=3~10^{11}~cm$ and $\rm V_C=3000~km/s$) we
obtain the light curve shown by the red solid line in Fig.~\ref{init_eclA}.
A larger size of the X-ray source would imply a correspondingly higher cloud velocity,
with the constraint that $\rm R_X/V_C(A) \sim 1~ksec$.

For cloud ``B'' the eclipse starts within an Earth occultation gap. In this case
we can only set an upper limit to the ratio $\rm R_X/V_C(B) < 3~ksec$, hence we
infer a lower limit on the cloud velocity of $\rm V_C(B)>1000~km/s$,
by assuming for $\rm R_X$ the smallest stable orbit for a maximally rotating
black hole.

We note that the lower limit on the 
velocity of cloud ``B'' is consistent with the average velocity
for BLR clouds inferred from the FWHM of the broad H$\beta$ line, i.e. $\rm 1900~km/s$
\citep{schulz99}, in agreement with the scenario where the eclipsing clouds
are part of the BLR, as discussed later on in Sect.~\ref{location}.

For cloud ``A'' we infer a velocity larger than expected for the
average BLR cloud, however not an unrealistic one given that the power law-like profile
of the broad lines \cite{nagao06}
implies a large fraction of clouds with velocity larger
than inferred from their FWHM.
Yet, if $\rm R_X\gg R_S$ then the clouds' velocity is unrealistically
large compared with the observed profile of the broad lines.

\subsection{The opening angle of the cometary tail}
\label{angle}
 
The evolution of the covering factor as a function of time provides constraints
on the geometry of the cloud tail. The model
used to fit the data (Fig.~\ref{model}) depends only on two parameters: the initial covering factor
($\sim 65\%$ both for cloud ``A'' and ``B'') and the quantity
$\rm R_X/(V_C \tan{\theta_{tail}})$,
where $\rm \theta_{tail}$ is the opening angle of the cloud tail.
In the case of cloud ``B'' (where the evolution of the covering factor is better
constrained) we obtain $\rm R_X/(V_C \tan{\theta_{tail}})\sim 35.8~ksec$. The
resulting fit to the data is shown with a dotted line in Fig.~\ref{eclipses}c. This value,
along with the upper limit on the ratio $\rm R_X/V_C(B) < 3~ksec$ inferred above,
implies an upper limit on the opening angle of the tail of $\rm \theta_{tail}(B)<4.8^\circ$.

In the case of cloud ``A'' the evolution of the covering is determined more poorly,
due to the larger errors. In this case we can only set an upper limit to the quantity
$\rm R_X/(V_C \tan{\theta_{tail}})>47.7~ksec$, whose corresponding fit to the
data is shown with a dotted line in Fig.~\ref{eclipses}c. However, as discussed above,
in this case we have a better constraint on the ratio $\rm R_X/V_C(A) \sim 1~ksec$,
hence we can obtain a stronger upper limit on the tail opening angle,
$\rm \theta_{tail}(A)<1.2^\circ$.

\subsection{The length of the cometary tail}
\label{length}

As shown in Fig.~\ref{eclipses}b, the column density $\rm N_{H,2}$
of each cloud drops rapidly
(the dotted lines represent a simple, empirical quadratic fit to the observed
evolution of $\rm N_{H,2}$).
The length of the tail is set observationally by our capability to disentangle
the cometary cloud absorption from the (constant)
global foreground absorber $\rm N_{H,1}$, which has a value of $\rm
7.5~10^{22}~cm^{-2}$ (Table~\ref{pars_comm}).
Through simulations we have obtained that our spectra cannot recover
a partial absorber with $\rm N_{H,2} < 2~10^{22}~cm^{-2}$
(i.e. a factor of $\sim$10 below the $\rm N_{H,2}$ in the head).
With this observational ``definition'', the time elapsed between the passage of the cloud head
and the time when $\rm N_{H,2}$ drops below this limiting value, along with the lower limit
on the cloud velocity inferred above, provides a lower limit on the length of the tail:
$\rm L_{tail}(A)>2~10^{13}~cm$ and $\rm L_{tail}(B)>10^{13}~cm$.

\section{How common are ``comets'' in AGNs?}

In this section we discuss whether the cometary geometry of the clouds observed by us
are peculiar to these observations, or to NGC~1365,
or whether they may be a more general property of the
absorbing clouds in other AGNs, not so far recognized so far in the data.

NGC~1365 may itself appear a peculiar case, since it is the AGN where the strongest
and most frequent $\rm N_H$ variations have been observed \citep[e.g.][]{risaliti05a,risaliti07,risaliti09a}.
This is partly due to the
fact that NGC~1365 is extremely bright in the X-ray band, which is a requirement to properly monitor
$\rm N_H$ variations even on short time scales. Some of the other nearby bright AGNs have indeed
displayed $\rm N_H$ variations on short time scales \citep{elvis04,puccetti07}.

However, among the other bright sources, which could
in principle allow the same investigation, NGC~1365 probably has a fortunate viewing angle.
If the obscuring clouds have a flattened distribution \citep{elvis00,risaliti02,maiolino01,gaskell07},
coplanar with the dusty circumnuclear absorber \citep{nenkova08}, then in most type 1 AGNs it is rare to
have a cloud transiting along the line of sight. On the other hand, in the classical type 2 AGNs, with
equatorial viewing angle, the number of clouds transiting along the line of sight is so large that
they make, on average, the absorption nearly constant. NGC~1365, which is a
type 1.8 AGNs \citep{schulz99},
probably has an intermediate viewing angle,
where there are a significant number of clouds transiting
the line of sight, but not large enough to average out the $\rm N_H$ variations.
Therefore, in this scenario, NGC~1365
is a particularly good laboratory to investigate clouds through their transits along
the line of sight, even though intrinsically it may have no
peculiar properties. An alternative possibility is that the density of absorbing clouds surrounding
different AGNs vary from object to object (in this scenario the difference between
type 2 and type 1 AGN
would be mostly due to the number of clouds rather than inclination), and
NGC~1365 may have the appropriate density of clouds that makes
 the probability of detecting eclipses high, but not large enough to average out
 $\rm N_H$ variations.

Assessing the detectability of the cometary geometry of the clouds in other objects similar to NGC1365,
as well as in other observations of NGC~1365 itself, is not simple.
Long ($\rm >100~ksec$), continuous observations, like the one reported here, are required to detect the
cometary tail in a transiting cloud. This kind of observation is performed rarely.
In most cases long integration times are obtained through several,
discontinuous short observations (a few $\rm 10~ksec$ each).
Even when such data are available the analysis
must be tuned to find these types of changes. This involves
the complex simultaneous fit over the various time intervals by leaving free both
covering factor and $\rm N_H$ for the variable absorber. Luck also plays a role: 1) the transiting
clouds must be moving fast to make the covering factor and $\rm N_H$ variations easy to
identify in terms of a cometary shape; 2) the variation of $\rm N_H$ and of covering factor
must be pronounced, else the two quantities would result in a degenerate fit;
3) if the head of the cloud does not eclipse the X-ray
source, and only the tail transits in front of the X-ray source, then the evolution of the light
curve and of the covering factor would appear much smoother, and would not be easy to ascribe to the cometary
tail.

To our knowledge, the only previous (nearly) continuous, moderately long observations of a bright AGN,
with variable obscuration events, are
the XMM long observations of NGC~1365 presented in \cite{risaliti09a},
\cite{risaliti09b} and \cite{turner08}.
The observation in  \cite{risaliti09a} does not show an ``eclipsing'' event, but rather
the uncovering of the X-ray source due to a ``hole'' in the clouds distribution, hence
it is not suitable to trace the geometry of the clouds. In the data of \cite{risaliti09b} the variation
of $\rm N_{H,2}$ and $\rm CF$ are much milder than observed in our data; as a consequence, in
\cite{risaliti09b} it was not possible to leave both parameters free in all time intervals,
since the two quantities would result degenerate. Therefore, in \cite{risaliti09b} only the
covering factor was left free to vary among the time intervals and details on the structure of the
eclipsing cloud could not be inferred. The inferred evolution of the CF during
cloud transit appears relatively symmetric, with no obvious evidence for a cometary shape. However,
if we had also kept $\rm N_{H,2}$ constant in our observation, then we would have obtained
a more symmetric variation of the covering factor: the decreasing $\rm N_{H,2}$ obtained
by us in the second
part of each eclipse (Fig.~\ref{eclipses}) would be interpreted (with a worse
$\chi^2$)
as a decreasing covering factor in a model where $\rm N_{H,2}$ is kept constant. Therefore,
the eclipse observed in the XMM observation may also be cometary-like, but the adopted model
may have prevented its identification. Yet, the light curve of the long XMM observation does
not show the sharp flux drop observed at the beginning of our eclipses ``A'' and ``B''.
This may be either due to the fact that the cloud in the XMM observation was moving
more slowly (hence the eclipse transition was much smoother in time), or to the fact
that the head of the cometary cloud did not pass in front of the X-ray source (hence
only the smoother occultation produced by its tail was observed), or more simply because
the cloud head was ``fuzzier''.

Similar considerations apply to the eclipse identified by \cite{turner08}, which was not
modelled in terms of both variable covering factor {\it and} variable $\rm N_H$, and whose
light curve does not show a sharp flux drop, as in \cite{risaliti09b}.

However, sharp drops of the X-ray flux in few ksec (similar
to those observed at the beginning of eclipses ``A'' and ``B''), have been observed
in previous observations of other AGNs. \cite{mckernan98} investigated in detail one such drop in an
ASCA observation of MCG-6-30-15, which they
ascribed to the occultation of the X-ray source by a ``symmetric'' cloud. However, they could not
investigate the spectral evolution of this occultation, which is required to
infer the presence of a cometary tail. By only using the light curve the beginning of
our eclipse ``A'' would be interpreted as an occultation by a nearly symmetric cloud; only the
investigation of the spectral evolution reveals the cometary shape.

Finally, we mention that our team has found evidence for cometary clouds also in another
source (Risaliti et al. 2010, submitt.) through the detailed temporal analysis of X-ray data.

Summarizing, the cometary shape could be a common property of the X-ray obscuring clouds in general.
However,
this feature is difficult to identify if the observations are not adequate or the
source not bright enough, so the occurrence rate is presently unknown.

\section{Cometary clouds as BLR clouds}
\label{location}

Several papers have discussed that rapid changes (days-hours) of the absorbing column density
in AGNs support a location of the associated absorbing clouds within the radius of the Broad Line Regions
\citep{elvis04,puccetti07,risaliti05a,risaliti07,risaliti09a,risaliti09b}. 
The rapidly variable absorption in our {\it Suzaku} data can also be ascribed to BLR clouds. In the following
we provide a more quantitative justification of this scenario.

We assume that the absorbing clouds are in Keplerian rotation around the supermassive black hole.
In the case of a spherical cloud crossing the line of sight (with the X-ray source smaller or comparable
to the projected size of the cloud), the distance $\rm D$ of the absorber from the black hole can be inferred
from the crossing time, traced by the absorption variation, through the equation \citep{risaliti02}
\begin{equation}
\rm D\sim 6~10^{26}~t^2_{cr}~n^2_{cl}~ N^{-2}_{H}~R_S~[cm]
\label{distance}
\end{equation}
where $\rm t_{cr}$ is the crossing time in units of ksec,
$\rm n_{cl}$ is the cloud density, in units of $\rm cm^{-3}$, and $\rm N_{H}$ its column density, in
units of $\rm cm^{-2}$.
In our case clouds are certainly not spherical. However, we can approximate
the dense ``head'' of the cometary clouds with hemispheres, as discussed above.
In this case $\rm t_{cr}$ is twice the transition interval at the beginning of the eclipses,
i.e. $\rm t_{cr}\sim 4~ksec$ (by taking the average of clouds A and B, Sect.~\ref{velocity}).
For the column density of the
heads we can assume the average of those observed at the beginning of eclipses A and B,
i.e. $\rm N_{H}\sim 1.5~10^{23}~cm^{-2}$. The gas density of the cloud can be roughly estimated by assuming
$\rm R_{head}\gtrsim R_S\sim 6~10^{11}~cm$ (as justified in Sect.~\ref{comets}),
hence $\rm n_{cl}=n_{head}\lesssim N_H/2R_{head}\sim 10^{11}~cm^{-3}$.
In this case Eq.~\ref{distance} gives $\rm D\lesssim 2~10^{15}~cm$.
This value is consistent with that inferred by \citep{risaliti09b} through an independent analysis of
other eclipsing events.

Although this is
an approximate estimate, given the various assumptions involved, it is within the radius of the BLR
inferred for NGC~1365. Indeed, by using the $\rm R_{BLR}-L_X$ relation given in \cite{kaspi05} and the
intrinsic X-ray luminosity inferred in \citep[][adapted to our new
slope]{risaliti09c},
$\rm L_{2-10keV}\sim 5~10^{42}~erg~s^{-1}$,
we infer a radius of the broad line region of $\rm R_{BLR}\sim 10^{16}~cm$. This result strongly
supports the scenario identifying the cometary clouds found in the {\it Suzaku} data with BLR clouds.
It should be noted that the BLR radius given by the relation in \cite{kaspi05} refers to the clouds
predominantly emitting H$\beta$. However, the radius of the BLR actually spans about an order of magnitude
as inferred by reverberation studies \citep{kaspi99},
with high ionization lines being emitted predominantly in the inner BLR while low ionization lines
predominantly coming from the outer region. The distance inferred for the absorbers
($\rm D\lesssim 2~10^{15}~cm$), relative to the BLR radius inferred for the H$\beta$ emitting clouds,
suggests that the absorbers observed by us are mostly associated with the high-ionization clouds
(e.g. those emitting CIV).

We can infer the mass of individual clouds based on the density inferred above. In particular the
``head'' of the clouds has a mass of
$\rm M_{head}\gtrsim n_{head}~m_H~\frac{4}{3}\pi R_{head}^3\sim 4~10^{-11}~M_{\odot}$ (here
we have assumed that the cloud head is a full sphere, whose leading hemisphere is probed
by the observed variations of $\rm N_H$ and $\rm CF$, while the trailing hemisphere is observationally confused
within the cometary tail).
For what concerns the tail,
given its small opening angle we can approximate
$\rm R_{tail}\sim R_{head}\sim R_X\gtrsim R_S = 6~10^{11}~cm$. We can infer the density of the tail
by assuming that the cross section of the tail is axi-symmetric relative to the
direction of motion, hence $\rm n_{tail}\sim N_{H,tail}/R_{tail}\lesssim 10^{10}~cm^{-3}$
(Fig.~\ref{eclipses}). Out to the length that we can detect trough our observations
(Figs.~\ref{eclipses}-\ref{model}) we infer a mass of the tail of
$\rm M_{tail}\gtrsim n_{tail}~m_H~\pi R_{tail}^2~L_{tail}\sim 10^{-10}~M_{\odot}$.

From the mass of individual clouds we can derive the mass of the BLR if the density of clouds
is known. The latter can be
inferred from the frequency of the eclipsing events.
Our and past observations of NGC~1365 display eclipses in $\sim$1/3
of the monitoring time,
implying that about one hundred of absorbing clouds must be present
in the volume, $\mathcal{V}$, probed by our line of sight toward
the X-ray emitting source in one year, within the region hosting the absorbers.
We can write
$\mathcal{V} \rm \sim 2 R_{head}~( V_C\cdot 1 yr)~ D \sim 5~10^{-13}~pc^3$, where we have assumed
$\rm V_C=2000~km~s^{-1}$ and $\rm D \sim 2~10^{15}~cm$ as representative of the clouds velocity and
of their distance to the center (as discussed above).
The inferred volume density of
the absorbers is therefore $\rm 100/\mathcal{V} \sim 2~10^{14}~
pc^{-3}$. If we assume that this density is uniform within the radius of the BLR, then
we obtain a total number of clouds of about $\rm \mathcal{N_C}\sim 3~10^7$.

The total mass of the BLR inferred from the properties of the absorbers is
$\rm M_{BLR}^{abs}\approx \mathcal{N_C}~(M_{head}+M_{tail}) \approx 4~10^{-3}~M_{\odot}$.
The latter value is five orders of magnitude lower than inferred from photoionization
models. Indeed, according to \cite{baldwin03}, the mass of the BLR
inferred from photoionization models scales as
$\rm M_{BLR}^{phot}\approx 2~10^{-44} (\nu L_{\nu})_{1450\AA} ~M_{\odot}$
\citep[note that we do not use the monocromatic luminosity as in ][]{baldwin03}.
Assuming $\rm (\nu L_{\nu})_{1450\AA}/L_{2-10keV}\sim 7$ \citep[][]{young10,richards06}, for NGC~1365 we obtain 
$\rm M_{BLR}^{phot}\approx 0.7 ~M_{\odot}$.
Accounting for such discrepancy is not easy.
One possibility is that the X-ray source is much larger than assumed by us ($\rm R_{head}\approx R_X\gg R_S$).
The inferred density of absorbers scales as $\rm \sim 1/R_X$, while the mass of individual clouds scales
as $\rm \sim R_X^2$, therefore the mass of the BLR inferred from the absorbers scales as $\rm \sim R_X$.
However, to obtain $\rm M_{BLR}^{abs}$ similar to $\rm M_{BLR}^{abs}$ would require $\rm \sim R_X=200~R_S$,
not easy to account for by standard models. Moreover, as discussed in sect.~\ref{velocity},
$\rm R_X\gg R_S$ would imply clouds' velocities unrealistically high relative to those inferred
from the broad lines profile.

Alternatively, a significant fraction of the BLR may not be
accounted for by the rapidly variable absorbers detected here. We may be biased towards the detection
of ``small clouds'', simply because the ``eclipses'' method
favors the detection of clouds with size similar to the X-ray source; larger clouds
may well exist, but may be more difficult to detect through $\rm N_H$ variations.
In particular, part of the ``constant'' absorber ($\rm N_{H,1}=7.5~10^{22}~cm^{-2}$) may be associated with much
larger clouds in the BLR (possibly located in the outer part),
whose $\rm N_H$ variability is only detected on much longer timescales
\citep[e.g. see ][]{risaliti09a,risaliti09b}.
The mass of such large clouds may account for the bulk of the mass of the BLR.

Yet another possibility is that the mass of the BLR clouds inferred by photoionization models may be
erroneously (over-)estimated. Indeed, such models assume that the photoionization is due to a nuclear
point-like source (the accretion disk). However, as discussed in the next section, part of the
photoionizing flux may be produced {\it locally} by the shock generated by the interaction between
the clouds and the diffuse intracloud medium. This is expected to significantly relax the requirements of
classical BLR photoionization models in terms of mass.

\section{Dynamics and fate of the cometary clouds}
\label{dynamics}

The differential velocity of the gas in the tail relative to the head ($\rm V_{head}-V_{tail}$)
is simply given by the sound speed of the gas in the cloud ($\rm \sim 10~km/s$), since behind the
head there is no hot intracloud medium (``cleaned'' by the passage
of the head).
nor other agents preventing the gas from expanding freely.
Essentially, the tail follows the head in its motion,
and lags behind only by a small fraction of the bulk velocity.

We can estimate the cloud mass-loss through the tail, which is given by
\begin{equation}
\label{mloss}
\rm \dot{M}_{head} = (V_{head}-V_{tail})~n_{tail}~m_H~\pi R^2_{tail}
\end{equation}
We obtain a mass-loss rate of about $\rm \dot{M}_{head} \approx 3~10^{-10}~M_{\odot}~yr^{-1}$. 
By comparison with the  mass of the clouds head inferred in the previous section
we infer a lifetime of the clouds of $\rm t_{life}\approx 2~months$, i.e. shorter than
their orbital period ($\rm 1 yr$).
A similar conclusion on the short lifetime of BLR clouds
was reached also by \cite{mathews87} based on the expected interaction of BLR clouds with the warm intracloud
medium (this issue will be discussed further in the next section).

The latter constraint has important implications.
The BLR in most
type 1 AGNs is observed to be present (and often nearly constant) for more
than 10-20 years. As a consequence, the much shorter lifetime of the BLR
clouds implies that they have
to be continuously replenished, and at a very small radius ($\rm <10^{16}~cm$). Direct infall
at such small radii is unlikely (inflow must occur through the accretion disk). The more likely
possibly is that BLR clouds are continuously produced by the accretion
disk, as suggested by the model in \cite{elvis00}.

It should be noted that $\rm t_{life}$ depends on our assumption
on $\rm R_X$, since $\rm M_{head}/ \dot{M}_{head}$ depends linearly on $\rm R_{head}\sim R_X$.
However, to make the lifetime of the absorbing clouds longer than several years (to account for the
observationally continuous presence and ``stability'' of the BLR) would require $\rm R_X\gg R_S$, which runs
into other physical problems discussed in sect.~\ref{location}.
Yet, within this context,
we note that the argument on the short timescale only applies to the ``small'' cometary clouds
detected through the short timescale $\rm N_H$ variability. The putative larger BLR clouds mentioned
in the previous section may not suffer the same timescale problem if they are much more massive.

The global mass loss of BLR comets is
$\rm \dot{M}_{tot} = \mathcal{N_C} \dot{M}_C \sim 10^{-2} ~M_{\odot}~yr^{-1}$.
Modelling the fate of such gas lost by the comets is very difficult and goes beyond the scope
of this paper. Here we only mention two possibilities. If the density drops significantly,
the effective ionization parameter may become high enough to make the gas thermally unstable and
heating up to $\rm 10^7~K$, hence contributing to the hot intracloud medium (see next section).
The gas lost by the cometary clouds may also be accelerated by the radiation pressure contributing to
a fraction of
the wind often observed in Seyfert nuclei, and whose outflow rate is generally estimated to be
around $\rm 0.1~M_{\odot}~yr^{-1}$ \citep{krongold07,crenshaw03,andrade10}.

\section{The nature of the cometary clouds}
\label{nat}

It is beyond the scope of the present work to define a detailed physical model
accounting for the nature of the
``cometary-like'' structure of the clouds. In this section we only try to identify the most 
plausible scenarios, by exploiting the constraints provided by the observations 
discussed in the previous sections.
As discussed above, these cometary clouds are most likely part of the BLR.
Therefore, any discussion on their nature is necessarily interlaced with the nature
of the BLR clouds in general, which is still matter of debate \citep[see ][for
a review]{netzer08}, although here we mostly focus on their cometary shape.

First, although we use the term ``comets'', these are not comets in the solar system sense,
i.e. their tail is not due to radiation pressure
on the gas. Indeed, the latter would make the tail elongated radially and yielding symmetric
variations of the $\rm N_H$ and of the covering factor, in contrast to what is observed.
Yet, as mentioned in sect.~\ref{comets} we cannot exclude a component of the tail elongated along
our line of sight.

In Sect.~\ref{location} we inferred that each cloud head must have
a gas density of about $\rm 10^{11}~cm^{-3}$, while
the density drops by about an order of magnitude along the tail.
The large density of the head may suggest that we are actually
seeing the (bloated) atmosphere of a star passing along the line
of sight. Models ascribing the BLR clouds to the atmosphere
of (bloated or evolved) stars have been proposed in the past
\citep{scoville88,alexander94,alexander97}
and it was also suggested that such stellar atmospheres may produce
contrails \citep{scoville95}, which may mimic
the ``tails'' observed in our clouds. However, as discussed in sect.~\ref{location},
the inferred density of the absorbing clouds ($\rm \sim 10^{14}~pc^{-3}$)
is by far larger than the stellar density in the center of galaxies.
In the light of the present X-ray 
observations, modelling the BLR structure in terms of stellar
atmospheres seems not to be a physically viable interpretation. 
The absorbing (and BLR) clouds are likely
purely gaseous (non self-gravitating) objects. Yet, we cannot exclude
that {\it some} of the eclipsing events are due to transiting stellar atmospheres.

Within the general framework of AGN central regions,  
these ``cometary-clouds'' should be moving through a diffuse, 
(relatively) hot intracloud medium (HIM),
as originally discussed in
\cite{krolik81}. This scenario has been subject
to revisions and objections and, at
present, the temperature of such diffuse medium is estimated to be
$\rm T_{HIM} \sim 10^7~K$ \citep{perry85,fabian86,collin88,netzer90,krolik99}.
This intracloud medium might well be in outflow, but it is highly unlikely
that such medium is in orbital motion around the central black hole, so that we 
can exclude that the absorbing clouds with the inferred velocities are comoving with this 
ambient medium (at least for what concerns the orbital component of their motion).
A simple estimate of the HIM sound 
speed shows that these cometary clouds must indeed be 
moving supersonically through the HIM. In fact, the sound speed in the HIM is
$\rm (c_s)_{HIM}\sim 290~(T_{HIM}/10^7)^{1/2}~km~s^{-1}$, while the inferred velocities of the 
``cometary'' clouds are significantly larger (Sect.~\ref{velocity}).
The inferred Mach numbers are $\rm M_A > 20$ and $\rm M_B > 7$
(recall that only lower limits on the clouds velocities
were obtained in Sect.~\ref{velocity}).
These imply Mach cones with opening angles ($\rm <1.5^{\circ}$ and
$\rm < 4^{\circ}$ for clouds A and B, respectively) consistent with the tails
opening angles inferred by us through the
variation of the covering factor.

The supersonic motion of dense clouds in a less dense medium has
been extensively studied in the past, although mostly
in very different astrophysical contexts \citep{falle02,pittard05a,pittard05b}.
Such theoretical studies
show that in general a sort of (``cometary'') tail, behind the front 
shock does indeed tend 
to form. In numerical simulations the specific result (shape, physical
properties, etc...) depends on the particular physical processes accounted for 
(for example, whether radiative cooling is included or not) and on the physical parameters, 
especially on the density contrast between unshocked ambient medium and
cloud material, and on the Mach number.
In any case, a system of front shocks (bow-shock like) at the head of the 
cloud, and a sort of Mach cone, prolonging the bow-shock at the rear of the 
cloud-head itself, must form.
The ``cometary tail'' is made up by the debris of trailing 
material lost by the cloud head, because of the action both of  Rayleigh-Taylor and
Kelvin-Helmholtz 
instabilities \citep{krolik81,mathews87} and of other possible 
disrupting mass loss mechanisms \citep[e.g.][and reference therein]{pittard05b}. Moreover,
as discussed above, the lack of a confining medium on the trailing side of the cloud head
naturally leads to gas loss at the sound speed ($\rm \sim 10 km/s$) due to free expansion,
even in absence of any hydrodynamical instability.
An elongated structure, trailing behind the cloud denser head, is therefore
expected by models, and also with a cometary-shape. However, a more
quantitative comparison
with our data is extremely complex, since it requires that such models are
modified to match the physical conditions observed in AGNs and in NGC~1365 in particular.

It is important to note that, as first pointed out by \cite{collin88}, the shock
generated by the supersonic motion of the BLR clouds into the HIM is expected to generate
UV and X-ray radiation that can contribute significantly
to the ionization {\it in situ} of the cloud itself.
This effect requires a substantial revision of the
classical photoionization models of the BLR, which assume
photoionization by a nuclear, point-like radiation source.

Finally, we mention that magnetic fields can provide an additional mechanism that could
elongate the circumnuclear clouds into a filamentary, and possibly cometary, shape. Strong
magnetic fields are expected to surround accreting black holes, and have been proposed
as an alternative mechanism to confine the BLR clouds \citep{rees87,ferland88}, with the effect of
making them elongated along the field. Inferring the actual shape of the clouds
expected to originate from such a mechanism is also very complex and beyond the scope of
this paper, a more detailed analysis of these effects is deferred to
a future paper. Here we only note that the past models of magnetically confined BLR clouds
do predict elongated structures, but do not obviously predict dense heads as those observed here.

\section{Summary}

We have presented a detailed spectral analysis of a long ($\rm \sim 300~ksec$), continuous
{\it Suzaku} observation of the Seyfert nucleus in the galaxy NGC~1365.
The spectrum shows evidence
for an absorption component that is variable in time, both in terms of column density and
in terms of covering factor, which is ascribed to clouds eclipsing the X-ray source.
This is the first time that a temporally
 resolved X-ray spectral analysis is able to break the degeneracy between the
 evolution of the column density ($\rm N_H$) and the covering factor (CF) of the X-ray
 absorber.

We identify two main eclipses.
The temporal evolution of each eclipse is far from being symmetrical. The initial occultation is
very rapid (within $\rm \sim 1~ksec$) and covering about 65\% of the source. Subsequently the covering
factor increases, but more slowly, reaching unity in about 50~ksec. The absorbing column density of the
cloud is about $\rm 10^{23}~cm^{-2}$ at the beginning of each eclipse and decreases
afterwards. These results are inconsistent with a spherical geometry for the absorbing clouds.
The most likely geometry compatible with the observations is an elongated, ``cometary'' shape, with
a dense ``head'' ($\rm n\sim 10^{11}~cm^{-3}$, consistent with that expected for the clouds
of the BLR) and with a dissolving and expanding tail.

The data allow us to quantitatively constrain
the geometry, dynamics and location of such cometary clouds.
The cometary clouds are probably located at a distance of about $\rm 2~10^{15}~cm$ from the nuclear
black hole, well within the estimated BLR radius ($\rm \sim 10^{16}~cm$), strongly supporting the
association of these absorbing systems with BLR clouds (at least with the inner, high-ionization ones).
The cometary clouds ``head'' must have a size comparable to the X-ray source and must be moving
with velocity higher than about 1000~km/s (consistent with the velocity expected for the BLR clouds).
The cometary tail must be longer
than a few times $\rm 10^{13}~cm$. The tail opening angle
must be very narrow, less than a few degrees, and consistent with the opening angle of the Mach cone
expected from
the supersonic motion of the cloud into the hot intracloud medium.

We suggest that such cometary clouds may be common to most AGNs, but have been difficult to
recognize in previous X-ray observations.

We estimate that the cloud ``head'' loses a significant fraction of its mass through the cometary
tail, which is expected to cause the total cloud destruction within a few months. If these clouds are
representative of most BLR clouds (or at least the high-ionization ones),
our result implies that the BLR region must be continuously
replenished with gas clouds, possibly from the accretion disk.

We briefly discussed the possible nature of such cometary
clouds. The most likely scenario is that the tail is made of gas lost by the cloud head through
hydrodynamical instabilities generated by its supersonic motion through the hot intracloud medium.

We can estimate the mass of the absorbing clouds ($\rm M_{cloud}\sim 10^{-10}~M_{\odot}$)
and their total number
within the central region ($\rm \mathcal{N_C}\sim 3~10^7$). The inferred total mass of the BLR
is about $\rm 4~10^{-3}~M_{\odot}$, which is two orders of magnitude lower than the BLR mass inferred
from photoionization models. The discrepancy may originate from a population of large, massive
BLR clouds not identified in our eclipsing studies. Alternatively, photoionization models may
overestimate the BLR mass. In particular, UV and X-ray radiation produced by the shocks generated
by the supersonic motion of the clouds may provide a local source of ionizing photons, not accounted
for by classical photoionization models that assume a central point-like radiation source.

\begin{acknowledgements}
We are grateful to A. Marconi and to C. Perola for useful discussions.
      This work was partially supported by INAF and by ASI.
	  
\end{acknowledgements}


\begin{thebibliography}{}

\bibitem[Alexander 
\& Netzer(1994)]{alexander94} Alexander, T., \& Netzer, H.\ 1994, \mnras, 270, 781

\bibitem[Alexander 
\& Netzer(1997)]{alexander97} Alexander, T., \& Netzer, H.\ 1997, \mnras, 284, 967

\bibitem[Andrade-Vel{\'a}zquez et al.(2010)]{andrade10} 
Andrade-Vel{\'a}zquez, M., Krongold, Y., Elvis, M., Nicastro, F., 
Brickhouse, N., Binette, L., Mathur, S., 
\& Jim{\'e}nez-Bail{\'o}n, E.\ 2010, \apj, 711, 888

\bibitem[Baldwin et al.(2003)]{baldwin03} Baldwin, J.~A., 
Ferland, G.~J., Korista, K.~T., Hamann, F., 
\& Dietrich, M.\ 2003, \apj, 582, 590

\bibitem[Collin-Souffrin et al.(1988)]{collin88} 
Collin-Souffrin, S., Dyson, J.~E., McDowell, J.~C., 
\& Perry, J.~J.\ 1988, \mnras, 232, 539

\bibitem[Crenshaw et 
al.(2003)]{crenshaw03} Crenshaw, D.~M., Kraemer, S.~B., \& George, I.~M.\ 2003, \araa, 41, 117 

\bibitem[Elvis(2000)]{elvis00} Elvis, M.\ 2000, \apj, 545, 63 

\bibitem[Elvis et al.(2004)]{elvis04} Elvis, M., Risaliti, G., 
Nicastro, F., Miller, J.~M., Fiore, F., 
\& Puccetti, S.\ 2004, \apjl, 615, L25

\bibitem[Fabian et al.(1986)]{fabian86} Fabian, A.~C., Guilbert, 
P.~W., Arnaud, K.~A., Shafer, R.~A., Tennant, A.~F., 
\& Ward, M.~J.\ 1986, \mnras, 218, 457

\bibitem[Falle et al.(2002)]{falle02} Falle, S.~A.~E.~G., 
Coker, R.~F., Pittard, J.~M., Dyson, J.~E., 
\& Hartquist, T.~W.\ 2002, \mnras, 329, 670

\bibitem[Ferland 
\& Rees(1988)]{ferland88} Ferland, G.~J., \& Rees, M.~J.\ 1988, \apj, 332, 141 

\bibitem[Gaskell et al.(2007)]{gaskell07} Gaskell, C.~M., Klimek, 
E.~S., \& Nazarova, L.~S.\ 2007, arXiv:0711.1025

\bibitem[Guainazzi et 
al.(2009)]{guainazzi09}
Guainazzi, M., Risaliti, G., Nucita, A., Wang, J.,
Bianchi, S., Soria, R., \& Zezas, A.\ 2009, \aap, 505, 589

\bibitem[Kaspi 
\& Netzer(1999)]{kaspi99} Kaspi, S., \& Netzer, H.\ 1999, \apj, 524, 71

\bibitem[Kaspi et al.(2005)]{kaspi05} Kaspi, S., Maoz, D., 
Netzer, H., Peterson, B.~M., Vestergaard, M., 
\& Jannuzi, B.~T.\ 2005, \apj, 629, 61

\bibitem[Koyama et al.(2007)]{koyama07} Koyama, K., et al.\ 
2007, \pasj, 59, 23

\bibitem[Krolik et al.(1981)]{krolik81} Krolik, J.~H., McKee, 
C.~F., \& Tarter, C.~B.\ 1981, \apj, 249, 422

\bibitem[Krolik(1999)]{krolik99} Krolik, J.~H.\ 1999, Active 
galactic nuclei : from the central black hole to the galactic environment 
/Julian H.~Krolik.~Princeton, N.~J.~: Princeton University Press, c1999.,

\bibitem[Krongold et al.(2007)]{krongold07} Krongold, Y., 
Nicastro, F., Elvis, M., Brickhouse, N., Binette, L., Mathur, S., 
\& Jim{\'e}nez-Bail{\'o}n, E.\ 2007, \apj, 659, 1022 

\bibitem[Maiolino et 
al.(2001)]{maiolino01} Maiolino, R., Salvati, M., Marconi, A., \& Antonucci, R.~R.~J.\ 2001, \aap, 375, 25

\bibitem[Mathews 
\& Ferland(1987)]{mathews87} Mathews, W.~G., \& Ferland, G.~J.\ 1987, \apj, 323, 456

\bibitem[McKernan 
\& Yaqoob(1998)]{mckernan98} McKernan, B., \& Yaqoob, T.\ 1998, \apjl, 501, L29

\bibitem[Mitsuda et al.(2007)]{mitsuda07} Mitsuda, K., et al.\ 
2007, \pasj, 59, 1

\bibitem[Nenkova et al.(2008)]{nenkova08} Nenkova, M., Sirocky, 
M.~M., Nikutta, R., Ivezi{\'c}, {\v Z}., 
\& Elitzur, M.\ 2008, \apj, 685, 160

\bibitem[Netzer(1990)]{netzer90} Netzer, H.\ 1990, Active 
Galactic Nuclei, 57

\bibitem[Netzer(2008)]{netzer08} Netzer, H.\ 2008, New Astronomy 
Review, 52, 257

\bibitem[Nagao et 
al.(2006)]{nagao06} Nagao, T., Marconi, A.,
\& Maiolino, R.\ 2006, \aap, 447, 157

\bibitem[Perry 
\& Dyson(1985)]{perry85} Perry, J.~J., \& Dyson, J.~E.\ 1985, \mnras, 213, 665

\bibitem[Pittard et al.(2005a)]{pittard05a} Pittard, J.~M., Dyson, 
J.~E., Falle, S.~A.~E.~G., \& Hartquist, T.~W.\ 2005, \mnras, 361, 1077

\bibitem[Pittard et al.(2005b)]{pittard05b} Pittard, J.~M., Dobson, M.~S., Durisen, R.~H.,
Dyson, J.~E., Hartquist, T.~W., \& O'Brien, J.~T.\ 2005, \aap, 438, 11

\bibitem[Puccetti et al.(2007)]{puccetti07} Puccetti, S., Fiore, 
F., Risaliti, G., Capalbi, M., Elvis, M., 
\& Nicastro, F.\ 2007, \mnras, 377, 607

\bibitem[Rees(1987)]{rees87} Rees, M.~J.\ 1987, \mnras, 228, 
47P 

\bibitem[Richards et al.(2006)]{richards06} Richards, G.~T., et 
al.\ 2006, \apjs, 166, 470

\bibitem[Risaliti et 
al.(2000)]{risaliti00} Risaliti, G., Maiolino, R.,
\& Bassani, L.\ 2000, \aap, 356, 33

\bibitem[Risaliti et al.(2002)]{risaliti02} Risaliti, G., Elvis, 
M., \& Nicastro, F.\ 2002, \apj, 571, 234

\bibitem[Risaliti et al.(2005a)]{risaliti05a} Risaliti, G., Elvis, 
M., Fabbiano, G., Baldi, A., \& Zezas, A.\ 2005, \apjl, 623, L93

\bibitem[Risaliti et al.(2005b)]{risaliti05b} Risaliti, G., Bianchi, 
S., Matt, G., Baldi, A., Elvis, M., Fabbiano, G., 
\& Zezas, A.\ 2005, \apjl, 630, L129

\bibitem[Risaliti et al.(2007)]{risaliti07} Risaliti, G., Elvis, 
M., Fabbiano, G., Baldi, A., Zezas, A., 
\& Salvati, M.\ 2007, \apjl, 659, L111

\bibitem[Risaliti et al.(2009a)]{risaliti09a} Risaliti, G., et al.\ 
2009, \mnras, 393, L1

\bibitem[Risaliti et al.(2009b)]{risaliti09b} Risaliti, G., et al.\ 
2009, \apj, 696, 160

\bibitem[Risaliti et al.(2009c)]{risaliti09c} Risaliti, G., et al.\ 
2009, \apjl, 705, L1

\bibitem[Scoville 
\& Norman(1988)]{scoville88} Scoville, N., \& Norman, C.\ 1988, \apj, 332, 163

\bibitem[Scoville 
\& Norman(1995)]{scoville95} Scoville, N., \& Norman, C.\ 1995, \apj, 451, 510

\bibitem[Schulz et al.(1999)]{schulz99}
Schulz, H., Komossa, S., Schmitz, C., \& M\"{u}cke, A. 1999, A\&A, 346, 
764

\bibitem[Soria et al.(2009)]{soria09} Soria, R., Risaliti, G., 
Elvis, M., Fabbiano, G., Bianchi, S., \& Kuncic, Z.\ 2009, \apj, 695, 1614

\bibitem[Turner et 
al.(2008)]{turner08} Turner, T.~J., Reeves, J.~N., Kraemer, S.~B., \& Miller, L.\ 2008, \aap, 483, 161

\bibitem[Turner et al.(2009)]{turner09} Turner, T.~J., Miller, 
L., Kraemer, S.~B., Reeves, J.~N., \& Pounds, K.~A.\ 2009, \apj, 698, 99

\bibitem[Wang et al.(2009)]{wang09} Wang, J., Fabbiano, G., 
Elvis, M., Risaliti, G., Mazzarella, J.~M., Howell, J.~H., 
\& Lord, S.\ 2009, \apj, 694, 718

\bibitem[Young et al.(2010)]{young10} Young, M., Elvis, M., 
\& Risaliti, G.\ 2010, \apj, 708, 1388



\end{thebibliography}
\end{document}